\documentstyle[12pt]{article}
\begin{document}
%\baselineskip=1.5\baselineskip
\begin{center}
{\Large Contrasting Quantum Cosmologies }\\
\bigskip
{\bf D.H. Coule}\\
\bigskip
Institute of Cosmology and Gravitation,\\ University of
Portsmouth, Mercantile House, Hampshire Terrace, Portsmouth PO1
2EG.\\
\bigskip

\begin{abstract}
We compare the recent loop quantum cosmology approach of Bojowald
and co-workers with earlier quantum cosmological schemes. Because
the weak-energy condition can now be violated at short distances,
and not necessarily with a high energy density,  a bounce from an
earlier collapsing phase might easier be implemented. However,
this approach could render flat space unstable to rapid expansion
or baby universe production; unless a Machian style principle can
be invoked. It also seems to require a flipping in the arrow of
time, and violates notions of unitarity, on passing through the
bounce. Preventing rapid oscillations in the wavefunction seems
incompatible with more general scalar-tensor gravity theories or
other classically accelerating solutions.

Other approaches such as ``creation from nothing''  or from some
quiescent state, static or time machine, are also  assessed on
grounds of naturalness and fine tuning.
\\

PACS numbers: 98.80.Qc

\end{abstract}
\end{center}
\newpage
{\bf 1.0 Introduction }

In recent work Bojowald and co-workers have applied notions of
loop quantum gravity to give an alternative quantum cosmological
description of the early universe - reviewed in [1]. This stems
from the idea that space comes in discrete packets i.e.
``granules'' or ``cells''. Such discreet quantum geometries can
evolve via Spin networks to smooth classical spaces - for general
review see [2].

At first sight such a scheme does not appear fruitful for
cosmology  since, in particular, flat or open
Friedmann-Robertson-Walker (FRW) models are infinite from the
start, so the granularity is never actually apparent. One can make
a cut to enclose the big bang within a finite volume as time
$t\rightarrow 0$, but then the matter or energy momentum becomes
infinite within such a domain [3,4] . By making a
compactification, at a scale $L$, this infinity can be alleviated
to some extent. But still the matter will diverge as the universe
is evolved back towards the initial singularity: likewise for
closed FRW models. Incidentally compactifying in this way
introduces possible vacuum polarization and Casimir like effects,
see e.g.[5,6]. Typically the energy density $\rho$ becomes
negative so violating the weak-energy condition. For a toroidal
model, one typically obtains a Casimir term, e.g.[5]
\begin{equation}
\rho=<T^0_0>=-\frac{\alpha}{L^4a^4}
\end{equation}
with $a$ the scale factor and the constant $\alpha$ depending on
the nature and number of matter fields present. More elaborate
twisted scalar fields can also be possible [7,5,6], which could be
detrimental if they are to drive chaotic style inflation [8].

 Interestingly, loop quantum gravity is also said to alter the
matter component such that the weak-energy condition is
effectively violated at short distances when the granularity of
space becomes significant. Recall that usually matter is diluted
for an expanding universe or remains constant for an exact de
Sitter solution. Likewise for the Casimir type component above.
But if the weak-energy condition is violated, together with a
positive energy density, it can instead grow even with expansion.
Such an example was called a ``whimper expanding to a big bang''
[9] or if relevant to the universe today, re-dubbed ``a big rip''
[10].

 By allowing the energy density to grow there
need not be a divergence in the energy density as the universe is
evolved back in time. One possible advantage is that now  a bounce
from an earlier collapsing phase might be implemented close to the
Planck length scale. Usually one needs to bounce before the Planck
energy density is surpassed, and also prevent ordinary matter from
dominating.

 This bounce case is apparently suggested in Bojowald's
loop quantum cosmology approach [1]. However this seem to have a
number of worrying if not serious consequences. We also contrast
this with more standard ideas in quantum cosmology such as the
Hartle-Hawking ``no boundary'' proposal [11],  ``creation from
nothing'' [12,13] or by starting from a possible quiescent state.
By concentrating on conceptual issues, without too much technical
distraction, we hope to see the strengths and weaknesses of the
various schemes. We suggest that for this formulism of loop
quantum cosmology, without a further type of Mach's principle, the
cells have apparently ``free rein'' and are not sufficiently
constrained by the global properties of the universe. We shall
exploit this weakness to show possible instability to runaway
expansion or baby universe creation.

{\bf 2.0 Loop cosmology and super-inflation}

 Bojowald and co-workers have suggested an inflationary mechanism
can be driven due to the discreteness of spacetime in loop quantum
cosmology [14,15]. Such a  mechanism violates the weak-energy
condition and so is more extreme than standard inflation. This
so-called super-inflation  is actually of pole-law form which has
a number of unsatisfactory properties [16].

 1) There is a second rapidly
collapsing deflationary solution  that occurs to the future of a
curvature singularity. You might object that standard de Sitter
also has a collapsing branch but that is to the past of the
expansionary phase and can be bypassed, or matched to an earlier
non-inflationary phase. But here  such a deflationary solution has
to be avoided and could intervene during the collapsing phase
preceding a classical singularity cf.[17] . In ref.[18] they
worked with this deflationary solution to produce a first
collapsing before later expanding cosmological model.

2) Such a pole-law inflation has a corresponding growing Hubble
parameter that produces a blue spectrum of perturbations [19].
Because the size of gravitational waves is given by the Hubble
parameter it should not be allowed to become larger than $\sim
10^{-5} m_{pl}$ [20].

 3) The definition of inflation is given as the
ratio of final to initial scale factor $a_f/a_i$ [14,15]. However
there is another requirement that the universe be sufficiently
large and produce mass density $\sim 10^{-30} gcm^{-3}$ today
[21]. This requires that $a_f$ be $\sim cm$ size, so requiring the
parameter $j$ [1,14,15] to be extremely large.

4) It is suggested that even if the pole-law inflation is not
itself sufficient, a second  potential driven inflation could be
produced [15,22]. This is because the friction term in the
Klein-Gordon equation can change sign and drive the field up a
scalar potential. However, if the kinetic term is always given by
the expression $\dot{\phi} \propto a^{12}$ (cf. eq. (10) in ref.
[22]) one can show that the potential term dominates over the
friction term and any growth in $\phi$ is negligibly small. But
instead of using this expression to determine the initial
$\dot{\phi}$ the authors of ref.[22]  use some argument based on
quantum uncertainty to give a correspondingly larger initial
$\dot{\phi}$ : this then does allow the friction term to dominate.
In my view their initial value of $\dot{\phi}\sim 10^{-5}$ is
actually not ``small'' and for the relevant parameters a value of
$ \sim 10^{-22}$ would not be unwarranted. I wish to satisfy the
various semi-classical equations from the start which seems more
reliable than starting with ``off-shell'' values. So it is not yet
entirely  clear just how robust such a scheme is for driving the
initial field up its potential hill.

Incidentally, in Euclidean space the friction term is also
switched in the corresponding Klein-Gordon equation. This
``anti-friction''  mechanism has been used previously in
conventional quantum cosmology to explain  a large initial field.

{\bf 2.1 Is Loop quantum cosmology unstable?}

For a massless scalar field the energy density $\rho\propto
a^{-6}$ is now modified at short distance such that $\rho\propto
a^n $ with $n>1$ [1,14,15,23]. At larger scales determined by the
parameter $j$ the energy density regains its standard behaviour.
The energy density now disappears as $a\rightarrow 0$ and so is
actually indistinguishable from flat space. But this suggests a
danger that actually any Planck sized region is now potentially
unstable to this inflationary expansion. One might try and reason
that for a Planck length region  to inflate it requires a negative
pressure that will be quickly equalized by the greater average
pressure of the universe outside. This was one of the reasons that
creating a ``universe in the lab'' is difficult because of a
pre-existing background metric [24]. But while such an
equalization is taking place there is the possibility of a quantum
tunnelling occurring to a new baby universe. This does not
supplant the original universe but disconnects forming a new
universe. In standard potential driven inflation such a scheme
requires one to produce a high energy density false vacuum that
then has a minuscule chance of tunnelling to produce a new
universe [25]. But now any Planck size region automatically could
make such a transition providing topology changes are not
forbidden on other grounds, see e.g.[26] for introduction to
topology issues. In the ``lab'' it required huge effort to violate
the strong-energy condition, but now the weak-energy condition is
continually being violated at short distances. One might try and
quantitatively calculate this enhancement but there is another
ambiguity: Bojowald has introduced an arbitrary compactification
scale for flat or open cosmological models. This seems a common
occurrence in loop cosmology e.g. [27]. Usually, these cases have
infinite action due to infinite size and are discounted, see e.g.
[28]. But now with a finite volume $V$ and energy density
decreasing with size such universes are not apparently suppressed
on action principles alone, $S=V\int a^n dt \rightarrow 0$ as
$a\rightarrow 0$. Again in the language of ref.[9] a ``whimper can
expand to a big bang''. In the closed case a forbidden region is
present so that the created universe must start with at least a
certain size cf.[15]. Placing this value beyond the weak-energy
violating region might help suppress the universe creation  effect
but this would introduce fine tuning.

We can also consider the creation of the original universe {\em ex
nihilo}. Now even in standard quantum cosmology it isn't clear why
universes are not still being  created around us. You can try and
argue that the forbidden region creates a barrier that to
observers within the existing universe suppresses further universe
creation e.g.[29]. But this barrier is either absent or reduced in
loop quantum cosmology and as we have argued might make universes
``too easily'' produced.  Admittedly, loop quantum cosmology has
rather suggested that the universe evolves not from ``nothing''
but from a previously collapsing phase [1,17]. A classical bounce
is anyway possible if the strong, or for the more general case,
weak-energy condition is violated, see e.g.[30].  We have already
reviewed some problems that classical bouncing universes have,
such as entropy production during the collapsing phase due to
perturbation growth [31]. The way that this is apparently overcome
with loop quantum cosmology appears a contradiction with the
generalized 2nd law of thermodynamics [32] and also unitarity
issues of quantum mechanics, see e.g.[33]. Instead of the energy
density growing
 on approaching the singularity it rather
 decreases so that degrees of freedom are
 being removed: so the arrow of time is actually reversed during the collapse.
 This goes against notions in black hole physics
 that information should not actually be destroyed. We previously
 criticized the cyclic ekpyrotic universe model [34] for requiring a
 similar entropy reduction scheme before the universe could be
 reset: a low-energy driven inflation alone does not alone achieve this task [31].

 Even if a bounce does proceed correctly one is still left with
 understanding some earlier  ``initial state'' even if this is now at
 a large classical scale. It might be possible to use the
 Hartle-Hawking type boundary condition to give an initially large
 universe [35], but this assumes Euclidean spacetime can also be
 allowed disconcertingly
 up to arbitrary large size [36]. It still would not apparently explain
 why entropy should reduce during the ensuing collapsing phase or
 remain negligibly small during the bouncing period.

{\bf 2.2 Loop quantum avoidance of singularity?}

 We now wish to compare some further aspects of loop quantum cosmology
 with the standard approach. As an archetypal example first
 consider the massless scalar field in a closed
 FRW universe, given by the Wheeler-DeWitt (WDW) equation, e.g.[37]. We
 follow our earlier presentation of this example [38],
 \begin{equation}
 \left (\frac{\partial}{\partial a^2}+\frac{p}{a}\frac{\partial}{\partial
 a}-\frac{1}{a^2}\frac{\partial}{\partial \phi^2}-a^2 \right ) \Psi(a,\phi)
 =0
 \end{equation}
 where $p$  represents part of the factor ordering ambiguity.

 The WDW equation can be separated to
 \begin{equation}
 \left ( a^2\frac{d^2}{da^2}+pa\frac{d}{da}+\nu^2-a^4  \right )\Psi(a)=0
 \end{equation}
 \begin{equation}
 \left (\frac{d^2}{d\phi^2}+\nu^2 \right ) \Psi(\phi)=0
 \end{equation}
 with $\nu$ the separation constant.\\
 The solution to these equations can be obtained using MAPLE [39],
 \begin{equation}
 \Psi(a) \sim a^{(1-p)/2} \left\{ \alpha J_{i\nu/2}(ia^2/2)+
 \beta Y_{i\nu/2}(ia^2/2) \right\}
 \end{equation}
 \begin{equation}
   \Psi(\phi) \sim \exp (i\nu\phi)
   \end{equation}
 where $J$ and $Y$ are Bessel functions (see e.g. [40])
 and each term has an
 associated arbitrary constant $\alpha,\beta$, which we can choose accordingly.

 First consider the limit $a\rightarrow 0$.
 Using the asymptote $J_{\mu}(z)\sim z^{\mu}$ as $z\rightarrow 0$
 enables the solution to be expressed as
 \begin{equation}
 \Psi(a)\sim a^{(1-p)/2} \exp (i\nu\ln a)
 \end{equation}
 There is a divergence as $a\rightarrow 0$ producing
  an infinite oscillation representing the classical singularity
 as the kinetic energy of the scalar field diverges. A similar divergence  also occurs
 for $\phi\rightarrow
 \infty$. In loop quantum cosmology this divergence is regulated
 as the discreteness of space alters the effective density [1].

But even as it stands, the solution can anyway be regularized by
integrating over the arbitrary separation constant. Now the
integral
\begin{equation}
\Psi(a,\phi)\equiv \Psi(a)\Psi(\phi)\sim \int  \exp \left
(i\nu[\ln a +\phi] \right )d\nu
\end{equation}
is of the form $\int \exp (ixt)dt$ which by means of the
Riemann-Lebesgue Lemma tends to zero as $x\rightarrow \infty$ (see
eg. ref.[41]). The wavefunction is now damped as $a\rightarrow 0$
or $\phi\rightarrow \infty$. There is another possible divergence
for factor ordering $p>1$ but we have assumed its coordinate
invariant value of unity [42]. It has also been suggested that in
the context of wormhole solutions these milder divergences due to
the factor ordering are not particularly serious [43]. They are
also present for flat empty space, so they conceivably anyway
should be renormalized away.  For large scale factor the
wavefunction eq.(8), behaves as $\sim \exp(-a^2/2)$ so indicating
asymptotically Euclidean space.\footnote{Provided the combination
$J+iY$ which equals the first Hankel function $H^{(1)}$ [40] is
chosen.}

Although loop quantum cosmology can also deal with the divergence
of the scale factor we do not see how it necessarily can deal with
the $\phi$ variable divergence. So although the $\Psi(a)$ part of
the solution is suitably regularized the actual solution
$\Psi(a,\phi)$ still can oscillate at arbitrary short ``pitch'',
due to the matter component. Only simple ``on shell'' perfect
fluid models allow the matter to be expressed in terms of the
scale factor. For the FRW case the scalar field has an extra
degree of freedom over a perfect fluid, cf.[44].

 To see this more explicitly consider the Brans-Dicke
model which
 is derived from the following action
\begin{equation}
S=\int d^4x\sqrt{g}\left ( \phi R-\frac{\omega}{\phi}(\partial _
{\mu}\phi)^2\right )\;\;.
\end{equation}
For stability in Lorentzian space one requires $\omega>-3/2$.

Using standard techniques, the corresponding WDW equation can be
obtained [38],
 \begin{equation}
 \left (a^2\frac{\partial}{\partial a^2}+ap\frac{\partial}{\partial
 a}-\frac{\phi^2}{3+2\omega}\left\{ \frac{\partial}
 {\partial \phi^2}+\frac{q}{\phi}\frac{\partial}{\partial \phi} \right\}
 -a^4 \right ) \Psi(a,\phi)
 =0
 \end{equation}
 where $p$  and $q$ represent part of the now two-factor ordering ambiguities.

 The WDW equation again can be separated but only the equation for
 $\Psi(\phi)$ differs from the previous case

 \begin{equation}
 \left (\phi^2 \frac{d^2}{d\phi^2}+q\phi\frac{d}{d\phi}+(3+2\omega)\nu^2 \right ) \Psi(\phi)=0
 \end{equation}
 with $\nu$ the separation constant.\\
 The solution is given by
 \begin{equation}
   \Psi(\phi) \sim \exp (i\sqrt{B}\nu\ln \phi)
   \end{equation}
   for $q=1$ and $B=(3+2\omega)$. So the oscillatory divergence now occurs for
   $\phi\rightarrow 0$ as well. Again one can integrate over the separation constant
   to produce a regular wave function
\begin{equation} \Psi(a,\phi)\equiv \Psi(a)\Psi(\phi)\sim \int \exp
\left (i\nu[\ln a +\ln \phi] \right )d\nu
\end{equation}
 Since  in the limit $\phi\rightarrow 0$  the Planck length $l_{p}=G^{1/2}\rightarrow
 \infty$, you might expect loop quantum  cosmology to also regulate this divergence. But in the opposite limit
 $\phi\rightarrow \infty$ there is also a divergence now at large distance beyond the Planck
 length. Although less severe than the previous divergence
 the ``pitch'' can still develop at arbitrarily short distance. We expect also rapid oscillatory wavefunctions more
 generally for
 scalar-tensor gravity models, including  non-minimally coupled scalar fields cf.
 [45]. Also higher order correction to the gravitational action
 typically correspond to additional scalar fields in the Einstein
 frame, see e.g.[46]

 One can see a similar behaviour in the WDW solution for a pure
 cosmological constant. The pitch of the solution gets
 increasingly shorter at large distance $a$ - see for example Fig.(4) in ref.[11]. This is simply
 because the universe keeps accelerating and so the ``velocity''
 $\dot{a}\rightarrow \infty$. We have remarked [31] that if loop
 quantum cosmology regulated such solutions it could prevent
 eternal inflation to the future. However, now this singularity avoidance
 mechanism might over-constrain such models and prevent wanted solutions. Bojowald mentions [47] that
 such examples are merely ``infrared problems'' and can be ignored since the local curvature is still small.
   But this distinction
 seems arbitrary, especially since the loop approach matches to flat space
  in the early universe, and not just when large energy densities are present.

{\bf 3.0  Standard Quantum Cosmology}

We have described how the massless scalar field example can be
regulated by integration over the separation constant. However, in
the common boundary conditions such as ``no boundary'' or
tunnelling ones this constant is simply zero, see e.g.[48]. So by
fiat the matter is  expunged and any possible oscillatory
divergences are simply absent. This allows any inflationary matter
present to become dominant  during the early stage of the
universe, so avoiding ambiguities found in classical measures for
inflation [49]. One might question whether this imposition is
reasonable and indeed it has been suggested that ``zero point''
fluctuations would alone alter this picture [50].\footnote{ One
could make a similar argument against loop quantum cosmology now
concerning the apparent absence of Casimir term, or vacuum
polarization, as $a\rightarrow 0$} But it does remove any
potential singularity due to stiff matter, contrary to the
impression of Bojowald that singularities couldn't be removed in
standard quantum cosmology cf.[17]. It is true however, that the
presence of strong-energy violating matter is a strict requirement
for these common boundary conditions otherwise no natural
Euclidean or forbidden region would be present.

The presence of a forbidden region allows both exponentially
growing and decaying solutions so there can be  big differences in
prediction, see e.g.[37]. The action of a de Sitter solution
driven by a scalar potential $V(\phi)$ is negative $\sim
-1/V(\phi)$ [51] and whether one should first take the modulus
alters the corresponding tunnelling rate $\sim \exp (-{\rm
Action})$ behaviour. There is some dispute whether the
Hartle-Hawking case $\sim \exp(1/V(\phi)$ gives sufficient
inflation [52]. This depends on whether you can include  energy
densities beyond the Planck value.

 We have earlier questioned whether
this analogy of treating the universe like $\alpha$ decay or a
Scanning Tunnelling microscope is sensible [53]. In these atomic
examples the various particles already exist. But now the
``particle'' is the universe itself coming into existence. Another
point is that if the barrier is removed you would expect to get a
stream of particles from the reservoir. This barrier can be
removed in flat or open cosmological models but is replaced by the
infinite action of any matter filling a now infinite universe
[28]. But if we compactify the space at arbitrary small size we
can reduce the action to an arbitrary small amount cf. [54] . In
this case the ``reservoir of universes'' can drain away. We have
suggested that the loop quantum cosmology approach is especially
susceptible to this dilemma but it could also be taken as a {\em
reductio ad absurdum} to standard quantum cosmology as well.
Although the loop case also seemingly allows baby universes to
easily break off from a pre-existing universes.

There is also the issue of including different topologies and
geometries for the tunnelling amplitude in the more general case.
Because the number of manifolds for the hyperbolic case can
approach infinity it can overwhelm the usual suppression factor
for the creation of a single universe with a set topology [55].
The ``average'' topology might be able to predict the spatial
homogeneity of the universe [56]. But again is this really
reasonable? It implies an initial state or  ``reservoir'' of all
infinite possible topologies that should be included in the
amplitude. There are now infinitely many ``particles'' one for
each possible topology and geometry. Neither is it clear why just
a single universe with an average topology results and not that
many universes each with different topology form together. Working
with closed models, and so fewer possible topologies, see e.g.
[57], could alleviate this problem but the notion of curvature
itself will anyway become hazy at the Planck scale.

Indeed the way that curvature  is treated as a constant is rather
unsatisfactory. In FRW models the actually   local characteristic
$k$ is taken to be globally constant. In more general metrics the
curvature can become a function also of time and space $k(t,x)$
cf. Stephani models e.g.[58]. For the  FRW model with perfect
fluid $p=(\gamma-1) \rho$ the WDW potential takes the form,
e.g.[44]
\begin{equation}
U=ka^2-Aa^{4-3\gamma}
\end{equation}
where the constant $A$ can be obtained from the relation
$\rho=A/a^{3\gamma}$. For a forbidden or Euclidean region at small
scale factor $a$  requires $U>0$ which requires $k=1$ and
violation of the strong-energy condition i.e. $0\leq \gamma<2/3$.
However, in a more general inhomogeneous model this behaviour can
be drastically altered. For example in a Stephani model the
corresponding WDW potential becomes cf. [59]
\begin{equation}
U=\beta a^n-Aa^{4-3\gamma}
\end{equation}
so for $n>2$ the forbidden region can be either narrowed or absent
entirely  even for closed models $\beta>0$ and when  the
strong-energy condition is being violated. This example is
symptomatic of what, more realistically, can be expected as the
Planck epoch is approached. The presence of forbidden regions that
play such a prominent role might not then actually be present,
even in closed models undergoing inflation. A slight complication
is that a negative energy density can also create a forbidden
region. For example a flat toroidal universe has a Casimir term
corresponding to a $\beta>0$ and $n=0$ term  in eq. (15) [13].

Even without a forbidden region some boundary conditions might be
adapted to purely Lorentzian metrics, although the underlying
principle is then often less prescriptive cf.[54]- where the
``outgoing only '' aspect of the Tunnelling boundary condition was
implemented in such cases.

 {\bf 4.0 Universe from a quiescent or static state}

We have spoken of the universe starting from nothing or by
bouncing from a previously collapsing phase. A third possibility
is that originally the universe was initially stuck in some
unchanging quiescent state. Perhaps involving the presence of
closed timelike curves (CTCs), see e.g.[26] for review. This is
closely related to introducing a topological identification scale
as in Misner space.

Starting with Misner space, Gott and Li [50,60] obtained a
self-consistent adapted Rindler vacuum state for a conformally
coupled scalar field that remains finite at the Cauchy horizon,
unlike for the Minkowski case [61]. They then conformally
transformed this state to give a suitable vacuum state for
multiply connected de Sitter space. Such a de Sitter space with
CTCs could be a suitable initial state for the universe. It only
has retarded solutions so giving an ``arrow of time'' and is a
state of low entropy, actually of zero temperature [50] .

However, in Misner space this state was only possible with
identification scale $b=2\pi$, or $b=2\pi r_0$ for the multiple de
Sitter case [50,60] . Such an exact value is itself inconsistent
with notions of quantum uncertainty. We are wary of claims that
such a multiply connected de Sitter state is stable especially
since the relevant time loop is approximately $\sim$ Planck time,
only a plausibility argument has so far been made [62].

The actual procedure of balancing a negative starting vacuum with
a Hawking radiation due to the periodicity to give an empty vacuum
state has possible difficulties. The calculation makes use of the
periodicity producing a thermal state [63] . Such a state is
required to be a many particle state with technically a suitably
large Fock space, see e.g.[6]. But by being close to the Planck
scale one starts reducing the number of allowed states due to
holography type arguments [64]. This will start preventing an
exact thermal state, as also is expected during the final stages
of black hole evaporation [65] or in Planck scale  de Sitter space
[51]. This mismatch could then result in some fluctuations still
being present in the vacuum instead of a pure empty state, so
destabilizing the CTC.

 Neither is
it clear that the $b$ value, or the corresponding de Sitter one,
remain independent of different matter couplings $\xi$ or
potentials $V(\phi)$. A more realistic combination of matter
sources still appears divergent at the Cauchy horizon [66],
although an improved {\em self-consistent} renormalization
procedure [67] in Euclidean space might help regulate some of
these other cases.

 Creating this state in any case seems rather contrived. Recall
that the Rindler vacuum of accelerating observers requires
``mirrors and absorbing stray radiation'', before we then make any
topological identification [68]. One would need some more general
reason why such an initial state was actually present. The
analogous zero temperature state for charged Black holes has
proved difficult to obtain on grounds of stability [69].

Instead of requiring CTCs one might just allow a static state with
time still evolving normally from say $-\infty$.  There is a
recent emergent model [70], an update of the
Eddington-Lema\^{i}tre model [9] that starts from an Einstein
static universe. Because this model has no forbidden region, and
requires a balance of ordinary matter and cosmological constant,
it again will  be prevented by the usual boundary conditions that
bias against the normal e.g. radiation matter component. Neither
do we think that maximizing the entropy is a more suitable
boundary condition since the entropy actually grows later during
the inflationary stage cf.[70]. The emergent model is however
geodesically complete to the past unlike the previous case with
multiply connected de Sitter space.

Also such a model also requires a mechanism to stabilize the
Einstein static phase to homogeneous perturbations. More general
inhomogeneous models might allow this. For example, by altering
the curvature dependence as in eq.(15) one could produce a stable
static universe with a now flat $U=0$ WDW potential; or perhaps,
at least prevent collapse to the origin by means of a repelling
potential $U>>0$ around the origin $a=0$, cf.[48].

On might also try to stabilize the Einstein static universe more
generally by surrounding the state entirely with forbidden or
Euclidean regions. For example if the sign of the WDW potential U
is flipped the corresponding Einstein static universe is
stabilized. Such a model requires $k;\Lambda;\rho;$ $\rightarrow
-k;-\Lambda;-\rho$, so now this is an open Anti-de Sitter with
negative radiation model. So violation of the weak-energy
condition is now required for such stability, such as might occur
in the Casimir effect cf. eq.(1). This could also be achieved
without altering the matter component by use of a signature
change, represented by the parameter $\epsilon$: $\epsilon =1$ for
usual Lorentzian space and $-1$ for Euclidean space [71]. In the
simplest case the corresponding WDW potential is altered
$U\rightarrow \epsilon U$ [44]. So if  for some reason
$\epsilon=-1$ the previous static universe is stabilized. Other
possible examples starting from different action principles are
also possible [72].

It has been suggested that oscillations of $\epsilon$ between the
two cases are constantly occurring but that the ``average'' is now
in the Lorentzian region [73]. One might imagine   instead a
preponderance of negative Euclidean values for $\epsilon$. This
might help stabilize a static model before for some reason the
sign changed and Lorentzian evolution then could proceed.

Despite these present difficulties the notion of finding  a
suitable quiescent state has some attraction. The difficulty, as
in the examples given, is why the state should survive for
semi-infinite times, but still have some slight instability that
causes the expansionary evolution to begin.

 {\bf 5.0  Conclusions}

 Although most work on quantum gravity is still being done within string
 theory, loop quantum gravity is also making progress. One
 advantage of loop quantum gravity over strings is that a
 background independent formulism might easier  be achieved [2]. One
 drawback however is that in GUT theories the various forces of
 nature should eventually unify. Therefore the present weak force of gravity
 should increase with energy scale to eventually coincide with
  the other forces of
 nature. The Planck length $G^{1/2}$ will correspondingly grow as
 the unification is achieved. But in loop gravity this aspect of
 ``running'' Planck length is
 apparently not incorporated at present. Incidentally, such a
 large initial Planck scale could  alter predictions for
 the initial state and any subsequent requirements of
 inflation.

 The idea of space being made of discrete quanta might introduce
 further conceptual problems. In an expanding model new cells have
 to be produced to fill in the gaps. But if we make analogy with
 cell division in living organisms, how are cells produced without
 error? Because presumably there is no analogy with DNA, there
 seems the need of providing ``scaffolding'' to force cells to have
 their correct form. One might try and claim the classical
 equations impose this by stricture, but if only a few cells are
 present the classical structure is still unformed. Constant
 quantum fluctuations at short distance have continually now to be
 kept in check.

 Because loop quantum cosmology allows the weak-energy condition
 to be violated it allows more variety than with more standard matter
 sources. Of course one might allow such phantom matter in
 standard quantum cosmology  but
 then there is no apparent scale where the effect can be turned
 off: the growing energy density will also eventually grow beyond
 Planck values that we some some confidence in describing.
  However, such properties are in danger of making flat space
 itself unstable to expansion, or to baby universe production.
 Bojowald has suggested to me that, in the above language, there
 now exists a scaffolding preventing such exotic behaviour due to the
 universe now obeying  the ``average'' classical description. We
 are suspicious how individual cells know about the average
 and so behave appropriately. Of course in the formulism this
 is innocuously hidden in the scale factor, which plays this ``non-local''
 messenger role. In standard cosmology space is never created in
 this way but is simply stretched like an elastic band by the
 scale factor. Even then gravitationally bound systems e.g.
 galaxies can drop out of the  expansionary global behaviour
 of the universe: so the scale factor never plays a universal
 messenger role to individual atoms. We seem instead, for loop cosmology, in need of
 a sort of ``generalized Mach's principle''e.g.[4,9], telling the individual granules how
 the universe actually is on average. A related concern [74] seems present
  in certain variable constant theories (see e.g. [75]), that also use the scale
  factor to determine, for example,  the varying speed of light.
  Maybe this adapted Mach's principle only needs to work up to some suitable classical size, a meter or so.
  Otherwise without such a constraint the instabilities I have
 suggested seem viable.

 The general idea that short distance oscillations in the
 wavefunction should be excluded due to granular effects might at first sight appear to be
 reasonable. However, such oscillatory behaviour is not readily
 confined to short distance in the scale factor {\em per se}. We can also question
 how Einstein's equations $G_{\mu\nu}\Leftrightarrow T_{\mu\nu}$
 have been
 used to interchange  between matter and geometry i.e.
 $\phi\leftrightarrow a$. Ideally this ``on shell'' relationship
 should be quantized and then not be used again. Care is also required if such relationships
  are used to remove degrees of freedom, that although irrelevant classically, can alter the full quantum
  theory cf.[44].

   In more general scalar-tensor gravity, or with higher order corrections
  to the gravitational action, this
   distinction between the geometry and matter is even more mixed up. The total solution can have arbitrary
   oscillations that cannot easily be confined or excluded by discreetness in the scale factor alone.
    Either this extra scalar
 should be excluded or is somehow itself unaffected  by the discreteness of space.
 Other standard models such as with a simple cosmological constant
 also get arbitrary short oscillation lengths corresponding to
 increasing kinetic energy. We have therefore suggested that this
 property of quantum gravity becoming important at short-distance is too
 general. One should require that also the energy-density be
 approaching large or even Planck values.

 The notion that a bounce could occur close to the Planck distance
 also seems to have unforseen consequences. Because
 matter is actually being diluted, even during contraction, the
 corresponding arrow of time always points away from the bounce
 point. Neither is this consistent with unitarity: that
 information should not readily be destroyed. Not only actual
 matter but the vacuum state itself must be adjusted. Otherwise,
 vacuum polarization effects (like Casimir and Conformal anomaly)
 would be expected to dominate cf.[76].

We have contrasted various common approaches to quantum cosmology
and admittedly none of them are without conceptual difficulties.
Of course some topics, particulary involving higher dimensions,
have not been considered here, but the issues are still rather
universal. Incidentally we have earlier criticized the use of
quantum cosmology in Brane type cosmologies on the grounds that
the initial bulk space is infinite, and Branes have large action
{\em ab initio}: neither of which is really amenable to a quantum
creation description [77]. Perhaps the bounce type scheme is
actually more appropriate, as attempted in the ekpyrotic universe
scheme, provided the singularity is suitably regulated.

The standard quantum cosmological schemes are found to have
certain {\em fragility} problems: like that forbidden regions are
strongly dependent on how the curvature behaves, or time machines
require extreme fine tuning. Analogies with atomic physics, such
as quantum tunnelling, are extrapolated to the universe as a
whole. Usually the boundary conditions have been developed
apparently with the sole aim of starting the universe in an
inflationary state. One then at least must start with matter
sources that could produce inflation. Although, there is still
some dispute whether this is achieved, it does help prevent
ambiguities in purely classical measures for the probability of
inflation. However, what preceded this inflationary state, and why
and how  it then evolved, is far from clear. Perhaps a better
explanation is still to be found within the realms of quantum
gravity. Whether a single quanta of spacetime - a modern version
of Lema\^{i}tre's ``primeval atom''- is involved or an entirely
new conceptual approach ( from M theory?) certainly remains a
fascinating  topic for the future.

{\bf Acknowledgement}\\ This work developed from interesting email
discussions with Martin Bojowald. I should also like to thank
William Hiscock for remarks on the renormalization issue for CTCs.

\newpage

{\bf References}\\
\begin{enumerate}
\item M. Bojowald and H.A. Morales-Tecotl, preprint
gr-qc/0306008.\\
M. Bojowald, Gen. Rel. and Grav. 35 (2003) p.1877.\\
M. Bojowald, preprint astro-ph/0309478.
\item A. Ashtekar, preprint math-ph/0202008 and references
therein;\\
for a simple introduction see also:\\ C. Rovelli, Phys. World 16
(2003) p.37\\
L. Smolin, ``Three roads to quantum gravity'' (Oxford University
Press: Oxford) 2000.
\item W. Rindler, Phys. Lett. A 276 (2000) p.52.
\item W. Rindler, ``Relativity: Special, General and
Cosmological'' (Oxford university Press: Oxford) 2001.
\item V.M. Mostepanenko and N.N. Trunov, ``The Casimir effect and
its applications'' (Oxford University Press: Oxford) 1997.
\item N.D. Birrell and P.C.W. Davies ``Quantum fields in curved
space'' (Cambridge University Press: Cambridge) 1982.
\item C.J. Isham, Proc. R. Soc. A 362 (1978) p.383.

\item A.D. Linde, Phys. Lett. B 129 (1983) p.177.
\item E. Harrison, ``Cosmology 2nd ed.'' (Cambridge University Press:
Cambridge) 2000
\item R.R. Caldwell, Phys. Lett. B 545 (2002) p.23.
\item J.B. Hartle and S.W. Hawking, Phys. Rev. D 28 (1983) p.2960.
\item A. Vilenkin, Phys. Rev. D 30 (1984) p.509.\\
A.D. Linde, Sov. Phys. JEPT 60 (1984) p.211.\\
V.A. Rubakov, Phys. Lett. B 148 (1984) p.280.
\item Y.B. Zeldovich
and A.A. Starobinsky, Sov. Astron. Lett. 10 (1984) p.135.
\item M. Bojowald, Phys. Rev. Lett. 89 (2002) p.261301.
\item M. Bojowald and K. Vandersloot, Phys. Rev. D 67 (2003)
p.124023.
\item D.H. Coule, Phys. Lett. B 450 (1999) p.48.
\item M. Bojowald, Phys. Rev. Lett. 86 (2001) p.5227.\\
M. Bojowald and F. Hinterleitner, Phys. Rev. D 66 (2002) p.104003.
\item F.G. Alvarenga and J.C. Fabris, Class. Quant. Grav. 12
(1995) p.L69.
\item S. Mollerach, S. Matarrese and F. Lucchin, Phys. Rev. D 50
(1994) p.4835.
\item V.A. Rubakov, M.V. Sazhin and A.V. Veryaskin, Phys. Lett. B
115 (1982) p.189.\\
L.F. Abbott and M.B. Wise, Nucl. Phys. B 244 (1984) p.541.
\item Y.B. Zeldovich, ``My Universe'' (Harwood Academic Press)
1992 p.95.
\item S. Tsujikawa, P. Singh and R. Maartens, preprint
astro-ph/0311015
\item M. Bojowald, Class. Quant. Grav. 19 (2002) p.5113.
\item E.H. Fahri and A.H. Guth, Phys. Lett. B 183 (1987) p.149.
\item E.H. Fahri, A.H. Guth and J. Guven, Nucl. Phys. B 339 (1990)
p.417.\\
W. Fischler, D. Morgan and J. Polchinski, Phys. Rev. D 42 (1990)
p.4042
\item M. Visser, ``Lorentzian Wormholes'' (AIP Press: New York)
1996.\\
see also F. Dowker in ref.[33]
\item S. Alexander, J. Malecki and L. Smolin, preprint
hep-th/0309045.
\item D. Atkatz and H. Pagels, Phys. Rev. D 25 (1982) p.2065.
\item R. Graham and P. Szepfalusy, Phys. Rev. D 42 (1990) p.2483.
\item C. Molina-Paris and M. Visser, Phys. Lett. B 455 (1999)
p.90.
\item D.H. Coule, Int. J. of Mod. Phys. D 12 (2003) p.963.
\item J.D. Bekenstein, Nuovo Cimento Lett. 4 (1972) p.737.\\
J.D. Bekenstein, Phys. Rev. D 7 (1973) p.2333.
\item L. Susskind, in ``The Future of Theoretical Physics and
Cosmology'' eds. G.W. Gibbons, E.P.S. Shellard and S.J. Rankin
(Cambridge University Press:Cambridge) 2003.
\item P.J. Steinhardt and N. Turok, Phys. Rev. D 65 (2002)
p.126003.
\item D. Green and W.G. Unruh, preprint gr-qc/0206068.
\item D.H. Coule, Mod. Phys. Lett. A 10 (1995) p.1989.
\item J.J. Halliwell, in ``Quantum Cosmology and Baby Universes''
eds. S. Coleman, J.B. Hartle, T. Piran and S. Weinberg (World
Scientific: Singapore) 1991.\\
D.L. Wiltshire, in ``Cosmology: the Physics of the Universe'' eds.
B. Robson, N. Visvanathon and W.S. Woodlcock ( World Scientific:
Singapore) 1996.
\item D.H. Coule, Mod. Phys. Lett. A 13 (1998) p.961.
\item Computer software, MAPLE V release 4.6, Waterloo Maple Inc.
1996.
\item M. Abramowitz and I.A. Stegun, ``Handbook of Mathematical
Functions'' (Dover Press) 1965.
\item C.M. Bender and S.A. Orszag, ``Advanced Mathematical Methods
for Scientists and Engineers'' (McGraw-Hill: ) 1984.
\item S.W. Hawking and D.N. Page, Nucl. Phys. B 264 (1986) p.185.
\item S.P. Kim, Phys. Rev. D 46 (1992) p.3403.
\item A. Carlini, D.H. Coule and D.M. Solomons, Mod. Phys. Lett. A
18 (1996) p.1453.
\item D.H. Coule, Class. Quant. Grav. 9 (1992) p.2353.\\
A.K. Sanyal, Int. J. Mod. Phys. A 10 (1995) p.2231.
\item G. Magnano and L.M. Sokolowski, Phys. Rev. D 50 (1994)
p.5039.
\item M. Bojowald, Class. Quant. Grav. 18 (2001) p.L109.
\item M.B. Miji\'{c}, M.S. Morris and W. Suen, Phys. Rev. D 39
(1989) p.1496.
\item G.W. Gibbons, S.W. Hawking and J.M. Stewart, Nucl. Phys. B
281 (1987) p.736.\\
D.N. Page, Phys. Rev. D 36 (1987) p.1607.\\
S.W. Hawking and D.N.Page, Nucl. Phys. B 298 (1988) p.789.\\
D.H. Coule, Class. Quant. Grav. 12 (1995) p.455.
\item J.R. Gott and Li-Xin Li, Phys. Rev D 58 (1998) p.023501.
\item G.W. Gibbons and S.W. Hawking, Phys. Rev. D 15 (1977)
p.2738.
\item A. Vilenkin, Phys. Rev. D 37 (1988) p.888.\\
D.N. Page, Phys. Rev. D 56 (1997) p.2065.
\item D.H. Coule, Phys. Rev. D 62 (2000) p.124010.
\item D.H. Coule and J. Martin, Phys. Rev. D 61 (2000) p.063501.
\item S. Carlip, Phys. Rev Lett. 79 (1998) p.4071.\\
S. Carlip, Class. Quant. Grav. 15 (1998) p.2629.
\item M. Anderson, S. Carlip, J.G. Ratcliffe, S. Surya and S.T.
Tschantz, preprint gr-qc/0310002.
\item G.W. Gibbons, in ref.[33]
\item Krasini\'{n}ski, ``Inhomogeneous Cosmological Models''
(Cambridge University Press: Cambridge) 1997.
\item J. Stelmach and I. Jakacka, Class. Quant. Grav. 18 (2001)
p.2643.
\item Li-Xin Li and J.R. Gott, Phys. Rev. Lett. 80 (1998) p.2980.
\item W.A. Hiscock and D.A. Konkowski, Phys. Rev. D 26 (1982)
p.1225.
\item P.F. Gonz\'{a}lez-D\'{i}az, Phys. Rev. D 59 (1999)
p.123513.
\item J.S. Dowker, Phys. Rev. D 18 (1978) p.1856.
\item G. 't Hooft, preprint gr-qc/9310026.\\
L. Susskind, J. Math. Phys. 36 (1995) p.6377.
\item S.W, Hawking, Commun. Math. Phys. 43 (1975) p.199.
\item W.A. Hiscock, preprint gr-qc/0009061.
\item Li-Xin Li, Phys. Rev. D 59 (1999) p.084016.
\item V.L. Ginzburg and V.P. Frolov, Usp. Fiz. Nauk 150 (1986) p.4.\\
L.P. Grishchuk, Y.B. Zeldovich and L.V. Rozhanski, Sov. Phys. JEPT
65 (1987) p.11.
\item P.R. Anderson, W.A. Hiscock and B.E. Taylor, Phys. Rev. Lett.
85 (2000) p.2438.
\item G.F.R. Ellis and R. Maartens, preprint gr-qc/0211082.
\item G.F.R. Ellis, A. Sumeruk, D.H. Coule and C. Hellaby, Class.
Quant. Grav. 9 (1992) p.1535.\\
G.F.R. Ellis, Gen. Rel. Grav. 24 (1992) p.1047.
\item F. Embacher, Phy. Rev. D 51 (1995) p.6764.
\item F. Embacher, Phys. Rev. D 52 (1995) p.2150.
\item D.H. Coule, Mod. Phys. Lett. A 14 (1999) p.2437.
\item J. Magueijo, Rep. Prog. Phys. 66 (2003) p.2025.
\item A.A. Grib and Y.V. Pavlov, preprint gr-qc/0206040.
\item D.H. Coule, Class. Quant. Grav. 18 (2001) p.4265.
\end{enumerate}
\end{document}